\documentclass[aps,twocolumn,groupedaddress]{revtex4}
\DeclareMathAlphabet{\zch}{OT1}{pzc}{m}{it}

\usepackage[]{endnotes}

\usepackage[]{color}
\usepackage[]{ascmac}
\usepackage[]{amsmath}
\usepackage[]{amssymb}
\usepackage[]{amsfonts}
\usepackage[]{latexsym}

\usepackage[]{graphicx}
\usepackage[]{color}
\usepackage{bm}

\newcommand{\eas}[0]{\begin{eqnarray*}}
\newcommand{\eae}[0]{\end{eqnarray*}}
\newcommand{\les}[0]{\begin{equation}}
\newcommand{\lee}[0]{\end{equation}}
\newcommand{\leas}[0]{\begin{eqnarray}}
\newcommand{\leae}[0]{\end{eqnarray}}

\newcommand{\mpic}[1]{}
\def\streda{{St$\check{{\rm r}}$eda} }

\begin{document}


\title{Polarization as a topological quantum number in graphene
}


\author{Hideo Aoki}
\email{aoki@cms.phys.s.u-tokyo.ac.jp}
\affiliation{
Department of Physics, University of Tokyo,
Hongo, Tokyo 113-0033, Japan
}
\author{Yasuhiro Hatsugai}%
\email{hatsugai@rhodia.ph.tsukuba.ac.jp}
\affiliation{
Institute of Physics, 
University of Tsukuba, Tennodai, Tsukuba, Ibaraki 305-8571, Japan}

\date{\today}


%
\begin{abstract}
Graphene, with its quantum Hall topological (Chern) number reflecting 
the massless Dirac particle, is shown to harbor yet another 
topological quantum number.  
This is obtained by combining St$\check{{\rm r}}$eda's 
general formula for the 
polarization associated with a second topological 
number in the Diophantine equation
for the Hofstadter problem, 
and the adiabatic continuity, earlier shown to 
exist between the square and 
honeycomb lattices by Hatsugai et al.
Specifically, we can regard, from the adiabatic continuity, 
graphene 
as a ``simulator" of square lattice with 
half flux quantum per unit cell, which implies that 
the polarization topological numbers in graphene in weak magnetic fields 
is 1/2 per Dirac cone for the energy region between 
the van Hove singularities, 
signifying a new quantum number characterizing graphene.
\end{abstract}


\maketitle

{\it Introduction}--- 
Characterization of quantum states
with topological numbers appears in various phenomenon 
in condensed-matter physics\cite{
Thouless82,Thouless98,Hatsugai04e,Hatsugai05-char,Hatsugai06a,
Hasan10,Qi11}.  
Topologically nontrivial states 
are characterized by nonzero topological numbers, which 
replace the order parameters in systems characterized by 
spontaneous symmetry breaking. 
A canonical model is the quantum Hall effect (QHE) 
with the Chern number characterizing the topological 
properties, 
where lattice structures (or periodic potentials) 
make the systematics of the topological numbers 
(known as Hofstadter's problem.\cite{Hof76}) versatile.  
The topological numbers in this problem are determined 
by a Diophantine equation, 
the so-called TKNN formula\cite{Thouless82}, 
which was also obtained by the 
Widom-\streda formula\cite{Widom82,Streda82}.  
Hofstadter's problem has been examined not only in condensed-matter 
systems but recently also in 
cold-atom systems with greater experimental 
controllability\cite{PhysRevLett.111.185301,PhysRevLett.111.185302}.  
A correspondence between bulk and edge topological properties 
(bulk-edge correspondence\cite{Hatsugai93b}) in the Hofstadter problem 
has in fact been confirmed in cold-atom systems\cite{Goldman12}.  
Lattice structure exerts an unusual effect on topological numbers, 
where a canonical example is 
honeycomb lattice as in graphene, where a QHE 
specific to massless Dirac 
particles appears\cite{Nov05,Zhang05,Hatsugai06gra,PhysofGra},  
indicating that topological numbers can be dramatically 
affected by the lattice structure.

Now, in the Diophantine equation, there exists actually 
a {\it second topological number}, which has long been known 
but its physical meaning was revealed only recently 
by \streda and coworkers, 
where the second topological number is shown to 
represent an electric polarization\cite{Streda06,Streda07,Streda08}.
These (i.e., graphene QHE and polarization 
quantum numbers) have motivated us to look into the 
following question: Does graphene harbor another  
topological number peculiar to a massless Dirac system 
besides the QHE number?  
We shall show that there exists indeed an intriguing polarization 
quantum number in graphene.  
To derive this we have fully exploited an {\it adiabatic continuity  
between the topological numbers for 
square and honeycomb lattices} as Hatsugai and coworkers 
have earlier shown\cite{Hatsugai06gra}, with which we can 
obtain a correspondence between the topological numbers for 
the two lattices.  With this we can obtain 
St$\check{{\rm r}}$eda's polarization topological numbers 
in graphene. 
An intriguing starting point is that the adiabatic continuity 
implies that graphene 
is a ``half-flux simulator" (an adiabatic realization of square lattice with 
half flux quantum per unit cell), whose consequence 
is that 
the polarization topological number in graphene in weak magnetic fields 
is 1/2 for a wide energy region (that encompasses the two 
van Hove singularities).  Thus this provides 
a novel topological quantum number characterizing graphene.

{\it Diophantine equation and topological numbers}--- 
Let us start with the
Hofstadter problem, i.e., non-interacting 
fermions in a two-dimensional, periodic lattice or potential in a uniform
magnetic field, for which the Hamiltonian in the tight-binding case reads 
$H = - \sum_{ij}  e^{i\theta_{ij} } c_i ^{\dagger}c_j$,
where we have taken the transfer energy as the unit of energy, 
and the Peierls phase $\theta_{ij}$ takes care of 
the magnetic flux $\phi=(2\pi)^{-1}\sum_{i,j} \theta_{ij}$ per unit cell in units of the flux quantum $\Phi_0\equiv h/e$.  
If we vary $\phi = p/q$ over rational values with 
$(p,q)$ mutually prime integers, we have a fractal energy 
spectrum (Hofstadter's butterfly). 
The  Diophantine equation for the topological quantum 
numbers\cite{Thouless82} is
\begin{eqnarray}
r = t_r p + s_r q \equiv t_r p \quad ({\rm mod}\, q),
\label{diophantine}
\end{eqnarray}
where $r$ labels the energy gaps from below, 
$t_r$ is the QHE topological (Chern) number, 
while $s_r$ is the number in question. 
For a given set of values of $(r,p,q)$ 
a pair of integers $(t_r,s_r)$ can be determined. 
For a square lattice with nearest-neighbor hopping, 
the Diophantine equation has a unique solution if 
one imposes 
$|t_r| \le q/2$.  
This constraint is perturbatively justified with an adiabatic argument. 
It is rather surprising that the integer $t_r$, which is determined by 
an algebraic (i.e. Diophantine) equation 
should coincide with a differential geometrical (topological Chern) 
number,
as confirmed by the Widom-\streda argument\cite{Widom82,Streda82}.   
The formula shows, via a Maxwell relation, that 
$\sigma _{yx} =
e(\partial n/\partial B) = 
(\partial  \rho/\partial B)$, 
where $n=N/A$ is the density of electron with 
$N$ the total number and $A$ the total area of the system.
With $\phi = A_0B/\Phi_0, r/q = nA_0$, 
where $A_0=a^2$ is the area of a unit cell,  we have 
$\partial   (r/q)/ \partial  \phi = (e/h) (\partial   n/\partial B)$, 
which reads, when combined with the Diophantine equation, 
$r/q = t_r \phi + s_r$, 
\begin{eqnarray*} 
\sigma _{yx} = e \frac {\partial n}{\partial B} = 
\frac{e^2}{h} t_r,
\end{eqnarray*}  
i.e., $t_r$ exactly coincides with the Chern number.  

The Diophantine equation reads in the original units as 
\begin{eqnarray}
n  = \left(\frac {\sigma _{yx}}{e^2/h}\right)\left(\frac {B}{h/e}\right) + \frac{s_r}{A_0},
\label{eq:Avss}
\end{eqnarray} 
or
\begin{eqnarray}
\rho' &\equiv & 
\rho-\delta \rho  = \sigma _{yx} B, 
\label{eq:rdio}
\end{eqnarray} 
where $\rho\equiv en$ is the charge density, and 
$\delta \rho \equiv (e/A_0)s_r$.
\streda and coworkers have shown that 
$\delta \rho \propto s_r$ is 
the electric polarization induced by the magnetic field 
as a quantum effect.\cite{stredaPRB50,Streda06,Streda07,Streda08}  
Indeed, Eq.(\ref{eq:Avss}) is expressed as 
$\frac{1}{A}\frac{\partial N}{\partial A_0}|_{B} = -(s_r/A_0^2)$, 
which is just Eq.(14) in \cite{stredaPRB50}.  

Physically, 
the Lorentz force acting on a particle of velocity $v_y$ is 
compensated by the induced electric field $E_x$ 
in Laughlin's cylyndrical geometry. 
The condition for the balance is $e v_y B =  e E_x$. 
We can then express the Hall current as a flow of the ``screened" 
charge density, $\rho'=\rho-\delta \rho$, as 
$I_y/L_x = \rho' v_y, V_x/L_x = E_x$,
where $L_x$ is the width of the cylinder and $V_x$ the Hall voltage.
Then the Hall conductivity is written as
\begin{eqnarray*}
\sigma _{yx} &=& \frac {I_y}{V_x} 
= \frac {\rho' v_y }{E_x} =  \frac {\rho'}{B}, 
\end{eqnarray*} 
which indeed implies
$\rho' = \sigma _{yx}B$, 
as consistent with the  Diophantine Eq.(\ref{eq:rdio}).

\begin{figure}[ht]
\begin{center}
\includegraphics[width=8.0cm]{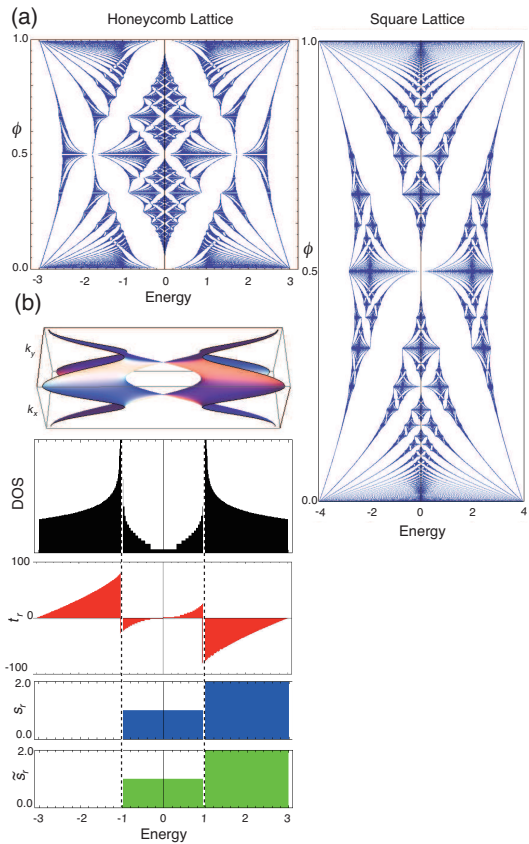}
\end{center}
\caption{(a) Hofstadter butterfly (one-particle energy spectrum vs 
magnetic field $\phi$) for the tight-binding model 
for honeycomb (left panel) or for square (right) lattices. 
To indicate the correspondence, $1/q \leftrightarrow 
1/2+1/2q$,  between the two, 
energy scales are doubled with a shift by 1/2.  
The butterfly for flux $\phi=p/q$ with all the prime 
$q\le 179 (1987)$ are plotted for honeycomb (square).  
Dashed line indicates the above correspondence. 
(b) For the honeycomb lattice, 
the band dispersion (with the 
energy as a horizontal axis), density of states $D(E)$, 
Chern number $t_r$, polarization topological number 
$s_r$ and screened polarization $\tilde{s}_r$ are plotted against energy 
for a weak $\phi=\delta (= 1/107$).  
Small gaps are neglected.
Dashed lines indicate van-Hove singularities.
}
\label{fig:adia}
\end{figure}

{\it Semiclassics around rational fluxes}--- 
Before addressing the adiabatic continuity, we need 
to 
look at the semiclassical behavior around rational fluxes 
for the polarization $s_r$.  
For the square lattice with the nearest-neighbor hopping, 
the Hofstadter's butterfly is as in Fig.\ref{fig:adia}, right 
panel, and 
the Chern number is given by the $t_r$ with $|t_r| \le q/2$. 
There is relatively a large gap for each simple fraction $P/Q$, 
and in its vicinity the semiclassical approximation should be appropriate,
so let us examine the polarization there.  
For this purpose one may define a reduced magnetic flux $\Delta \phi$ as
a small deviation from a simple flux $\Phi=P/Q$ as 
$\phi = p/q= P/Q + \Delta\phi$. 
We can then regard the Hofstadter problem at $\phi$
as composed of the {\em effective } Landau
levels formed by the reduced magnetic flux $\Delta \phi$.
The reduced magnetic field is naturally defined as
$\tilde{B} \equiv (\Phi_0\Delta \phi)/A_0$, which implies,
with the Diophantine equation (\ref{eq:rdio}) in 
the original units, that 
the reduced polarization $\delta \tilde{\rho}$ 
is 
\begin{eqnarray*}
\rho &=& \sigma _{yx}B + \delta \rho
\equiv \sigma _{yx}\tilde{B} + \delta \tilde{\rho}.
\end{eqnarray*} 
This defines the reduced polarization quantum number, 
\begin{eqnarray}
\tilde{s}_r &\equiv& \left(\frac{A_0}{e}\right) \delta \tilde{\rho}
= 
s_r + \sigma _{yx} \left(\frac{\Phi_0}{e}\frac{P}{Q}\right) 
\nonumber \\
&=& s_r+t_r\frac {P}{Q} = \frac {r}{q} - t_r \Delta \phi,
\label{eq:reduced-s}
\end{eqnarray} 
where we have used Eq.(\ref{diophantine}).

\begin{figure}
\begin{center}
\includegraphics[width=8.0cm]{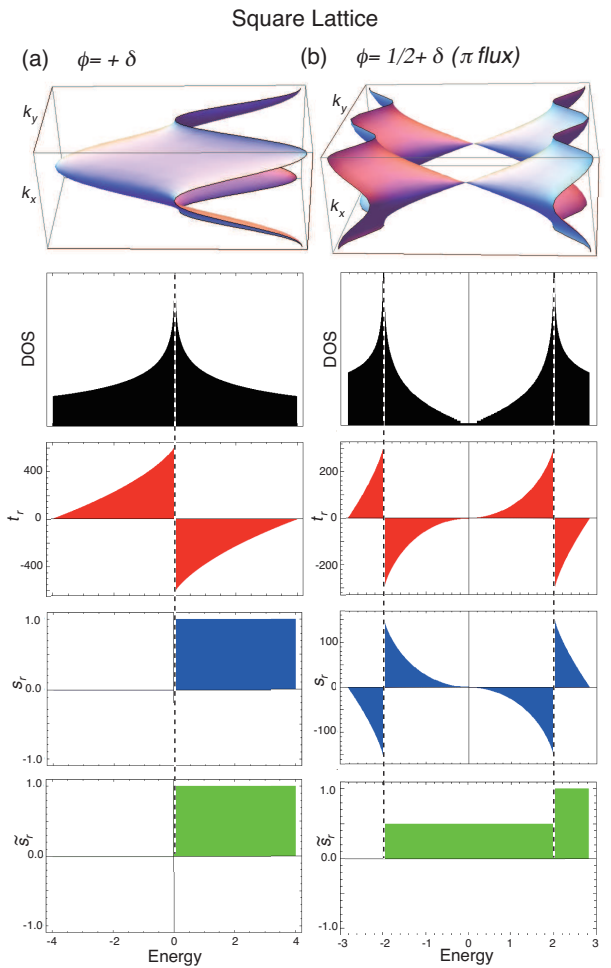}
\caption{
For the square lattice, the band dispersion (with the 
energy as a horizontal axis), density of states $D(E)$, 
Chern number $t_r$, polarization topological number 
$s_r$ and screened polarization $\tilde{s}_r$ are plotted against energy 
for a small $\phi=\delta$ (left panel) or around the 
$\pi$ flux, $\phi=1/2 + \delta$ (right) with $\delta=1/1223$.  
Small gaps are neglected.
Dashed lines indicate van-Hove singularities. 
}
\label{fig:case-0}
\end{center}
\end{figure}

We can illustrate this for three cases, $\Phi=0,1/2$ and $1$.  
Near $\Phi=0$, 
the standard Landau levels for a two-dimensional electron gas are realized for 
$\phi=1/q$.  
For the less than half-filled case (electron side; $r<q/2$),
we have 
$t_r =  r$, 
which trivially implies 
$\tilde{s}_r = s_r= (A_0/e) \delta\rho=0$.
On the hole side ($r>q/2$) 
the solution of the Diophantine equation is given by
$t_r = q-r$, which implies
$\tilde{s}_r = s_r=(A_0/e) \delta\rho= 1$.  
Namely, $\tilde{s}_r$ against the Fermi energy $E_F$ is given as
$\tilde{s}(E_F) = 0 (E_F<0), 1 (E_F>0)$, 
where $\tilde{s}_r$ is specified by the Fermi energy $E_F$ in the $r$-th gap
as $\tilde{s}(E_F) =\tilde{s}_r$\cite{comm}. 
Hence $\tilde{s}(E_F)$ is a simple step function 
with the step situated at $E=0$, square lattice's van Hove singularity in the
band dispersion in zero magnetic field. 
A numerical result for weak magnetic fields is shown in Fig.\ref{fig:case-0},
which confirms the analytical arguments.
Near $\Phi=1$, 
with $\phi
=(q-1)/q$  ($\Delta \phi=-1/q$), 
we can show that 
$\tilde{s}_r$ is given by the same function as in the above 
case for near $\Phi=0$, 
which is consistent with the periodicity of the Hofstadter problem.
Note that the original $s_r$ itself does not satisfy 
this periodicity,  which implies the reduced polarization 
$\delta \tilde{\rho}$ is more physical.

Now we come to the case in question, $\Phi=1/2$ ($\pi$-flux).  
As shown in \cite{Hatsugai06gra}, if we want to  adiabatically relate 
a honeycomb lattice with a flux $1/q^\prime$ 
to a square lattice, 
we can consider a square lattice with a 
flux $\Delta \phi=1/2q ^\prime $ on top of $\pi$ flux 
per unit cell, i.e.,
\begin{eqnarray*}
\phi &=& \frac {p}{q}= \frac{1}{2}+\Delta \phi=\frac {q ^\prime +1}{2 q ^\prime },
\end{eqnarray*} 
where $q^\prime +1$ is assumed to be prime with $2 q ^\prime$.  
As shown in \cite{Hatsugai06gra}, 
every step in the Chern number  $t_r$ against $E_F$ 
has a height of 2 everywhere except at the van Hove singularities of the $\pi$-flux band at $\pm 2\sqrt{2}$.  
The energy spectrum of the  Hofstadter problem
in the $\Delta \phi\to 0$ limit is composed of 
touching two bands (see Fig.\ref{fig:adia}).
Then near the band bottom of the $\pi$-flux bands below the van Hove energy
$-2\sqrt{2}$, we can put 
$r=2r^\prime$ ($r^\prime=1,2,\cdots, q ^\prime/2 $), 
for which the Diophantine equation, 
$2r^\prime \equiv (q ^\prime +1)t_{2r^\prime} 
\; ({\rm mod}\; 2 q ^\prime,\ |t_{2 r ^\prime }|< q^\prime)$, 
has a solution $t_{2r^\prime}=2r^\prime$, which implies 
$s_{2r^\prime} = [2r^\prime-t_{2r^\prime}(q ^\prime +1)]/2q ^\prime =-r ^\prime$. 
Then the reduced polarization quantum number is trivially
$\tilde{s} _{2 r ^\prime } = - r ^\prime + 2 r ^\prime(1/2)  =0$.
If we turn to the $r$-th gap on the hole side 
above the van Hove energy at $2\sqrt{2}$, 
we can put $r=2 (q ^\prime - r ^\prime) $, ($r ^\prime =1,\cdots, q ^\prime /2$), for which the Diophantine equation,
$2 (q ^\prime - r ^\prime) 
\equiv (q ^\prime +1) t_{2(q ^\prime - r ^\prime )}\: ({\rm mod}\ 2 q ^\prime )$
 has a solution 
$t_{2 (q ^\prime -r ^\prime) }=- 2 r ^\prime$, giving
$s_{2(q ^\prime-r^\prime)} = [2( q^\prime -r^\prime)-t_{2(q ^\prime -r^\prime)}(q ^\prime +1)]/2q ^\prime =r ^\prime +1$,
and the reduced polarization quantum number, 
$\tilde{s} _{2(q ^\prime - r ^\prime) } =  r ^\prime +1- 2 r ^\prime(1/2) =1$.

In the region of interest (for Dirac electrons 
residing between the van Hove singularities, $-2\sqrt{2}<E_F<2\sqrt{2}$) 
it is convenient to introduce 
the usual Landau index, $N (=0,\pm 1,\pm 2,\cdots)$, 
with $N=0$ corresponding to the level at $E=0$.  
As depicted in Fig.\ref{fig:adia}, there are $2q^\prime $ Landau bands, 
since each level is composed of two (with a tiny gap 
not visible in Fig.\ref{fig:adia}), so that we have 
$r = q ^\prime +1+2N$ ($N=-q ^\prime /2,\cdots-1,0,1,\cdots, q ^\prime /2-1$). 
The Diophantine equation, 
\begin{eqnarray*}
q ^\prime +1+2N &\equiv & t_r (q ^\prime +1)\; ({\rm mod}\ 2 q ^\prime ),\; 
|t_r|< q ^\prime 
\end{eqnarray*} 
then has a solution 
$t_r = 2(N+1/2)$, 
which is the Chern number for the Dirac fermions with doubling. 
The polarization is 
$s_r = [r-(2N+1)(q ^\prime +1)]/2 q ^\prime =-N$, 
and the reduced polarization quantum number becomes
\begin{eqnarray*}
\tilde{s} _r &=& -N+(2N+1)\cdot \frac {1}{2} =\frac {1}{2} .
\end{eqnarray*} 
So we end up with a key result,
\begin{eqnarray}
\tilde{s}(E_F) &=& 
\left\{
\begin{array}{lc}
0   &(E_F<-2\sqrt{2}) \\
\frac {1}{2}    &(-2\sqrt{2}<E_F<2\sqrt{2}) \\
1    &(2\sqrt{2}<E_F)
\end{array}
\right.
\label{eq:nearpi}
\end{eqnarray} 
In Fig.\ref{fig:case-0}(b) we have numerically calculated the Chern number $t_r$, 
the St$\check{{\rm r}}$eda's polarization $s_r$ and the reduced polarization 
$\tilde{s}_r$ against the Fermi energy 
for $\Phi=1/2$.  The result 
indeed confirms the analytical result, eq.(\ref{eq:nearpi}).

We have also calculated $t_r,s_r$ and $\tilde{s}_r$ 
for a series of rational fluxes, e.g. 
for $\phi_0=1/5, 2/5$ etc, 
and the results are rather surprising in that the reduced polarization 
around a rational flux $p/q$ is given quite simply as
\begin{eqnarray}
\tilde{s}(E_F) =  \frac n{ q},
\quad E_{\rm vH}^n < E_F< E_{\rm vH} ^{n+1},
\quad n=0,1,\cdots, q
\end{eqnarray} 
where $E{\rm vH}^n $ 
is the energy of van Hove singularities 
in the $n$-th Hofstadter band from below.  
 It signifies a {\em topological character of the reduced polarization
quantum number} $\tilde{s}_r$ in that only the (van Hove) 
singularities can 
change the sequence of $\tilde{s}_r$.

{\it Graphene}--- 
In a honeycomb lattice in a flux 
$\phi_{\rm h} =1/q $ 
per hexagon, 
a finding in Ref.\cite{Hatsugai06gra} 
is that the large gaps 
in graphene remain open in an adiabatic 
continuity between square and honeycomb lattices 
where the diagonal transfer $t'$  for each hexagon is 
changed from 1 to 0.  
Further surprise is that 
this holds not only around the zero energy, but all the way up to 
the van Hove singularity, $E^{1,2}_{\rm vH} \simeq \pm1$, for honeycomb lattice.
Since the adiabatic mapping between 
honeycomb and square lattices is given as 
\begin{eqnarray*}
&&{\rm honeycomb:}\;\; \phi_{\rm h} = 1/q\\ 
&&\leftrightarrow {\rm square:}\;\; \phi_{\rm s} =1/2+\phi_{\rm h}/2=(q+1)/2q,
\end{eqnarray*} 
i.e., a honeycomb lattice in weak 
magnetic fields translates into a square lattice around {\it half flux}, 
as depicted in Fig.\ref{fig:adia}, we can directly apply the 
above result for the square lattice around half flux. 
We have only to note that, 
similar to the $\pi$ flux case of the square lattice, 
the gap is  rewritten  in terms of the  Landau 
index, $N$, this time for graphene Landau levels (with 
$N=0$ labeling the level at $E=0$), which is related to $r$ as
$r = q+1+2N$.
By the adiabatic continuity for the Chern number, 
we have
$t_r = 2N+1$, 
which reproduces the graphene QHE number as doubled Dirac cone contributions 
($2(N+1/2)$). 
Then the polarization quantum number is 
given as
\begin{eqnarray*}
s_r = 1.
\end{eqnarray*} 
Below the lower van Hove energy, the Hall conductance is simply $t_r=r$ ($r<q/2$), 
that is, $s_r=0$.  
Above the upper van Hove energy, we have $t_r=-(2q-r)$ $(r>3q/2)$, which 
implies 
$s_r=2$.
Thus the polarization quantum number of the graphene against the Fermi energy is\begin{eqnarray*}
s = \tilde{s} &=& 
\left\{
\begin{array}{lc}
0  & (E_F<E_{\rm vH}^1 )  \\
1  & (E_{\rm vH}^1<E_F<E_{\rm vH}^2 )   \\
2  & (E_{\rm vH}^2<E_F)
\end{array}
\right.
\end{eqnarray*} 

Namely, since the unit cell area in honeycomb is twice 
that in $\pi$-flux square lattice, $\tilde{s}$ is doubled, which 
is consistent with the contribution of 1/2 per Dirac cone. 
Thus graphene can indeed be considered as the ``half-flux simulator" of
the Hofstadter problem. 
This is the key result for graphene.  
We have numerically calculated the polarization in Fig.\ref{fig:adia} (bottom 
left panel), 
which confirms this formula.  Since we take a weak magnetic field for 
graphene there, the reduced polarization 
coincides with the polarization itself 
($\tilde{s}_r =  s_r$).  

In general, however, $\tilde{s}_r$ and $s_r$  can 
deviate from each other as in square lattice even  for graphene.  
Near a generic rational flux $\phi_{\rm h}= P/Q$, one can define the 
reduced polarization quantum number as
$\tilde{s}_r = s_r +t_r (P/Q)$, 
which takes fractional values in general. 
For instance, 
the reduced polarization for graphene near the
$1/2$ flux resembles the case for the square lattice near the
$1/4$ flux (see the supplemental material).

To summarize, 
we have shown that graphene harbors, in addition 
to the quantum Hall topological number, another 
topological quantum number as the electric polarization.  
An interesting future problem is how to experimentally 
observe this, which may be 
possible if the electron density $n$ is independently 
measures, since the other topological number has to do with the 
difference between $\sigma _{yx} B$ and $\rho=en$.  

We wish to thank Pavel \streda for illuminating discussions.  
HA is also indebted to Mikito Koshino for discussions 
in an early stage of the present 
work.  The work has been supported in part by Grants-in-Aid for Scientific Research Nos. 23340112(HA, YH), 25107005(HA, YH), 25610101(YH), and 23540460(YH) from JSPS. 

\vfill


\end{document}